\documentclass[12pt,a4paper,final]{iopart}

\usepackage{iopams} 
\expandafter\let\csname equation*\endcsname\relax
\expandafter\let\csname endequation*\endcsname\relaxvariance
\usepackage{amsmath}
\usepackage{graphicx}
\usepackage{cite}
\usepackage[breaklinks=true,colorlinks=true,linkcolor=blue,urlcolor=blue,citecolor=blue]{hyperref}
\begin{document}

\title[Exact analysis of gate noise effects ]{Exact analysis of gate noise effects on non-adiabatic transformations of spin-orbit qubits}

\author{Lara Ul\v{c}akar$^1$ and Anton Ram\v{s}ak$^{1,2}$}
\address{$^1$ J. Stefan Institute, Ljubljana, Slovenia\\$^2$ Faculty of mathematics and physics, University of Ljubljana, Ljubljana, Slovenia}

\vspace{10pt}
\begin{indented}
\item[]3 May 2017
\end{indented}
\begin{abstract}
We considered various types of potential noise in gates controlling non-adiabatic holonomic transformations of spin-qubits in one and two dimensional systems with the Rashba interaction. It is shown how exact results can be derived for deviations of spin rotation angle and fidelity of the qubit transformation after a completed transformation. Errors in initial values of gate potentials and time-dependent drivings are considered and exact results for white gate noise  are  derived and analysed in detail. It is demonstrated how the drivings can be tuned to optimise the final fidelity of the transformation and to minimise the variances of qubit transformations.
\end{abstract}


\section{Introduction}  

The new branch of electronics, spintronics, has been the object of intense activity over the past decade since it promises  enhanced performance with smaller power consumption compared with classical electronics \cite{wolf01}. Spintronics has potential for realising the fundamental building blocks of a quantum computer via electron spin qubits. Implementation of such qubits is relatively simple in gated semiconductor devices based on quantum dots and quantum wires \cite{hanson07}. Qubit manipulation may be  achieved by rotating the electron's spin  by the application of an external magnetic field \cite{Koopens06}. However, this is unwieldy and not easily confined and controlled in small regions occupied by qubits. The main challenge is therefore how to accurately manipulate the spin of a single electron without using an external magnetic field.

A possible solution is to make use of the spin-orbit interaction (SOI). In semiconductor heterostructures there are two types of spin-orbit interaction, the Dresselhaus interaction \cite{dresselhaus55} due to bulk inversion asymmetry of a crystal, and the Rashba  interaction \cite{bychkov84} which is a consequence of structural inversion asymmetry of the confining potential of the two-dimensional electron gas. 
In spintronic devices the latter is particularly suitable for qubit manipulation since it can be tuned  locally via electrostatic gates. Furthermore, since the Rashba interaction couples electron's spin to its orbital motion, 
qubit spin rotation can be performed by adiabatic spatial translation of a quantum dot containing a single electron, for a distance of the order of the spin-orbit length \cite{stepanenko04,flindt06,coish06,sanjose08,golovach10,bednarek08}. In one-dimensional quantum systems also electric-field-induced resonance can manipulate electron spin \cite{fan16}, tunnel-coupled spin qubits can be driven by ac fields \cite{gomez12}, 
and most recently it was shown that time dependent Rashba interaction in a quantum wire can contribute to the  rotation of electron spin \cite{pavlowski16,pavlowski16b}. Experimentally such systems with the ability of controlling electrons
have been realised  in InSb \cite{nadjperge12}, InAs \cite{nadjperge10,fasth05,fasth07} and Ge \cite{shin12} quantum wires. Recently this type of qubit manipulation  has been generalised to non-adiabatic quantum dot motions due to external time-dependent potentials \cite{cadez13}, thus opening up the possibility of much faster spin-qubit transformations. 

The simplest non-adiabatic qubit manipulation with exact analytical solution is achieved by translating a qubit in one dimension in the presence of constant Rashba interaction  \cite{cadez13}. A drawback of such manipulation is that after the transformation, the qubit is trapped in a displaced quantum dot. This deficiency can be remedied by applying a time-dependent electric field which produces a time-dependent Rashba coupling \cite{nitta97,liang12}.
The qubit manipulation then consists of first displacing the quantum dot, followed by changing the Rashba coupling, then returning the quantum dot to the original spatial position and finally tuning the Rashba coupling to its initial value. Such a system thus represents a one-dimensional spatial motion in a two-dimensional parametric space - spanned by the position of the quantum dot and the strength of the Rashba coupling.
For quantum dots with harmonic confining potential the exact analytical solution is known for various quantum phases including 
non-adiabatic non-Abelian Anandan phase \cite{anandan88} which opens the possibility of exact qubit transformations  \cite{cadez14}. However, the transformations are limited to cases of rotations with fixed axis.
Most recently this limitation posed by fixed axis of spin rotation was also eliminated in a quantum ring structure where full coverage of the Bloch sphere is possible \cite{kregar16,kregar16b}. 

Exact solutions for all three methods of qubit manipulation also simplifies the analysis of possible effects of environment which result in decoherence and relaxation of the qubit's state. This arises due to fluctuating electric fields, caused by the piezoelectric phonons and conduction electrons in the circuit \cite{sanjose08,sanjose06}
due to ionized dopant nuclei in a heterostructure \cite{huang13} or the hyperfine interaction with the nuclei \cite{echeveria13}. In molecular systems with phonon assisted potential barriers phonon-mediated instabilities could introduce noise in the confining potentials \cite{mravlje06,mravlje08} and in non-adiabatic qubit transformations the effects are also related to the speed of the moving quantum dot \cite{cadez14}. Electrons could be carried also by surface acoustic waves, where additional noise could be introduced by the electron-electron interaction \cite{giavaras06,jefferson06}.

In this paper we concentrate on the analysis of errors of such qubit transformations and decoherence. Essential for a correct transformation is the precise application of external electric fields via various top gates. An important consideration in the practical implementation of this scheme is the effect of random fluctuations in both the time-dependent SOI and the QD motion as well as the influence of errors in the initialisation of qubit states. The paper is organised as follows. After the introduction, in Section 2 we introduce the model and show exact solutions of the time-dependent Schr\" odinger equation. In Section 3 errors in spin-qubit transformations in one- and two-dimensional parametric space are analysed. Exact expressions are given for white noise in electric potentials and an explicit example is presented. Next in Section 4 exact results for fidelity due to the white noise are derived and Section 5 is devoted to the summary and conclusion. 

\section{Model}

We consider an electron in a quantum wire confined in a harmonic trap \cite{cadez13,cadez14}. The center of the potential (one-dimensional quantum dot),  $\xi(t)$, can be arbitrarily translated along the wire by means of time dependent external electric fields. Spin-orbit Rashba interaction between electric field and electron spin couples with orbital motion, resulting in the
following Hamiltonian
\begin{equation}\label{H}
H(t)=\frac{p^{2}}{2m^{*}}I+\frac{m^{*}\omega^{2}}{2}[x-\xi(t)]^{2}I+\alpha(t)p\mathbf{n}\boldsymbol{\cdot\sigma},
\end{equation}
where $m^{*}$ is the electron effective mass, $\omega$ is the frequency
of the harmonic trap, $\alpha(t)$ is the strength of spin-orbit interaction, possibly time 
dependent due to appropriate time dependent external electric fields.  The spin rotation axis $\mathbf{n}$ is fixed and depends 
on the crystal structure of the quasi-one-dimensional material used and 
the direction of the applied electric field \cite{nadjperge12}. $\boldsymbol{\sigma}$
and $I$ are Pauli spin matrices and unity operator in spin space, respectively. Exact solution of
the time dependent Schr{\"o}dinger equation corresponding to the Hamiltonian equation~\eqref{H} is given by
\begin{equation}\label{psi}
|\Psi_{ms}(t)\rangle=e^{-i\omega_{m}t}\mathcal{U}^{\dagger}(t)|\psi_{m}(x)\rangle|\chi_{s}\rangle,
\end{equation}
\begin{equation}
\mathcal{U}^{\dagger}(t)=\mathcal{A}_{\alpha}\mathcal{X}_{\xi},
\end{equation}
\begin{equation}
\mathcal{A}_{\alpha}=e^{-i[(\phi_{\alpha}(t)+m^{*}\dot{a}_{c}(t)a_{c}(t)/\omega^{2})I+\phi_A(t)\mathbf{n}\cdot\boldsymbol{\sigma}/2]}
e^{-i\dot{a}_{c}(t)p\mathbf{n}\cdot\boldsymbol{\sigma}/\omega^2}
e^{-im^{*}a_{c}(t)x\mathbf{n}\cdot\boldsymbol{\sigma}},
\end{equation}
\begin{equation}
\mathcal{X}_{\xi}=e^{-i\phi_{\xi}(t)I}e^{im^{*}[x-x_{c}(t)]\dot{x}_{c}(t)I}e^{-ix_{c}(t)pI}.
\end{equation}
Here $\psi_{m}(x)$ represents the $m$-th eigenstate of a harmonic oscillator with eigenenergy $\omega_{m}=(m+1/2)\omega$ and $|\chi_{s}\rangle$ is spinor of the electron in the eigenbasis of operator $\sigma_z$.
Unitary transformations $\mathcal{A}_{\alpha}$ 
and $\mathcal{X}_{\xi}$ transform the system into the "moving frame" of SOI
and position, respectively, and therefore $\mathcal{U}^{\dagger}(t)$ transforms the Hamiltonian equation~\eqref{H}
into a simple time independent harmonic oscillator Hamiltonian. 
The phase $\phi_{\xi}(t)=-\int_{0}^{t}L_{\xi}(t')\mathrm{d}t'$ is the coordinate action
integral, with $L_{\xi}(t)=m^{*}\dot{x}_{c}^{2}(t)/2-m^{*}\omega^{2}[x_{c}(t)-\xi(t)]^{2}/2$
being the Lagrange function of a driven harmonic oscillator and $x_{c}(t)$ is the
solution to the equation of motion of a classical driven oscillator
\begin{equation}\label{eq:xoscillator}
\ddot{x}_c(t)+\omega^{2}x_{c}(t)=\omega^{2}\xi(t).
\end{equation}
Another phase factor is the SOI action integral phase 
$\phi_{\alpha}(t)=-\int_{0}^{t}L_{\alpha}(t')\mathrm{d}t'$,
where $L_{\alpha}(t)=m^{*}\dot{a}_{c}^{2}(t)/2-m^{*}a_{c}^{2}(t)/2+m^{*}a_{c}(t)\alpha(t)$
is the Lagrange function of another driven oscillator, satisfying
\begin{equation}
\ddot{a}_{c}(t)+\omega^{2}a_{c}(t)=\omega^{2}\alpha(t).
\end{equation}
Spin-qubits are rotated around $\mathbf{n}$ by two "dynamical" terms proportional to operators $a_{c}(t)x$, $\dot{a}_{c}(t)p$,  and
by the angle $\phi_A(t)=-2m^{*}\int_{0}^{t}\dot{a}_{c}(t')\xi(t')\mathrm{d}t'$,  the Anandan phase for the case of cyclic motions \cite{cadez14}.

\section{Spin-qubit transformations}
\subsection{One-dimensional parametric space}

First we consider a special case of constant SOI,  $\alpha(t)=\alpha_0$, which
means that the parameter space is one-dimensional (1D). The exact solution, equation~\eqref{psi} is completely determined by the classical response of the oscillator, equation~\eqref{eq:xoscillator}, which makes exact analysis of the qubit transformation very simple. For example,
if  the electron is initially in the $m$-th excited state of $H(0)$, the spin is rotated
around $\mathbf{n}$ for angle,
\begin{equation}
\phi(T)=2m^{*}\alpha_0 x_{c}(T),
\end{equation}
where $T$ is the transformation time  \cite{cadez13}. When the driving $\xi(t)$ is chosen to give full spin-flip $\phi(T)=\pi$, the final displacement of the electron is $x_{c}(T)=\pi/(2m^{*}\alpha_0)$  and no residual angle oscillations are present. This is fulfilled exactly when the final state of the electron is in the $m-$th eigenstate of $H(T)$, that is when $x_c (T)=\xi(T)$ and  $\dot x_{c}(T)=0$.

In qubit transformations of this type it is essential to control precisely the initial state and driving electric fields. 
Noise in fields of gate electrodes is reflected in fields
which translate the trap potential minimum and is consequently manifested as
noise in the initial position $\delta x_{c}(0)$, initial velocity $\delta \dot{x}_{c}(0)$
and driving function $\delta\xi(t)$. This produces noise in $x_{c}(T)$ which further
induces noise in $\phi(T)$. Using the exact solution, equation~\eqref{psi} it is straightforward to analyse the noise in the transformation angle $\phi$, which is dispersed by some probability distribution, given by a change of variables when the probability density function of $x_{c}$ is known.

We assume that errors in initial position and velocity may be described by normal
distributions with variances  $\sigma_{x0}^{2}$ and $\sigma_{\dot{x}0}^{2}$, respectively,
\begin{equation}
\frac{\mathrm{d}P}{\mathrm{d}x_{c}(0)}=\frac{1}{\sqrt{2\pi}\sigma_{x0}}e^{-x_{c}(0)^{2}/2\sigma_{x0}^{2}},\qquad
\frac{\mathrm{d}P}{\mathrm{d}\dot{x}_{c}(0)}=\frac{1}{\sqrt{2\pi}\sigma_{\dot{x}0}}e^{-\dot{x}_{c}(0)^2/2\sigma_{\dot{x}0}^{2}}.
\end{equation}
The driving function $\xi(t)=\xi^{0}(t)+\delta\xi(t)$ consists of ideal driving part without noise, 
$\xi^{0}(t)$ with superimposed  stochastic part with vanishing mean  $\langle\delta\xi(t)\rangle=0$ and
characterised by the time autocorrelation function $\langle\delta\xi(t')\delta\xi(t'')\rangle$. We consider here coloured noise, in particular the Ornstein-Uhlenbeck process \cite{wang45,masoliver92,meinrichs93}
with exponential correlations  $\langle\delta\xi(t')\delta\xi(t'')\rangle={{\sigma_{\xi}^2} \over {2\tau_\xi} }e^{ |t'-t''|/\tau_\xi}$ with noise intensity $\sigma^2_\xi$ and correlation time $\tau_\xi$. 

A general solution of equation~\eqref{eq:xoscillator} $x_{c}(t)$ is given by
\begin{equation}\label{eq:harmoicsolution}
x_{c}(t)=x_{c}(0)\cos \omega t+\dot{x}_{c}(0)\frac{1}{\omega}\sin \omega t+\omega\int_{0}^{t}\sin[\omega(t-t')]\xi(t')d t',
\end{equation}
where all three terms are stochastic, independent and normally
distributed variables. Their sum is also normally distributed
with variance equal to the sum of variances of all variables \cite{feller71}. The variance
of the first two terms is obtained by the change of variables formula
while the variance corresponding to the third term is evaluated as equal-times autocorrelation function \cite{uo30},
\begin{equation}
\sigma_x^2(t)=\omega^2\lim_{\Delta t \to0} \langle\int_{0}^{t}\sin[\omega(t-t')]\delta\xi(t')d t' \int_{0}^{t+\Delta t}\sin[\omega(t-t'')]\delta\xi(t'')d t'' \rangle.
\end{equation}
For the Ornstein-Uhlenbeck noise considered here the integrals can be evaluated exactly and the final result is that the angle of spin is distributed normally with the time dependent variance
\begin{equation}\label{1d}
\sigma_{\phi}^2(t)=(2m^{*}\alpha_0)^2\left(\sigma_{x0}^{2}\cos^{2}\omega t+\sigma_{\dot{x}0}^{2}{{\sin^{2}\omega t}\over{\omega^2}}+\sigma_x^2(t)\right),
\end{equation}
with $\sigma_x^2(t)=\frac{1}{4}\omega\sigma_{\xi}^{2}\left(2\omega t-{{\sin{2\omega t}}}\right)+{\cal O}(\tau_\xi)$, where only the short correlation time ($\tau_\xi \to 0$) contributions -- corresponding to the white noise in driving -- are explicitly shown here. The first two terms in equation~\eqref{1d}  are limited by the precision of the initial conditions while the third contribution, related to the noise in driving, diverges at large $\omega t$, since the Lorentzian noise power spectrum $ \sigma^2_\xi/[1+(2 \pi f \tau_\xi)^2]$ considered here consists of different driving frequencies including the resonant value $\omega=2\pi f$, resulting in the asymptotic response $\sigma_{x}^2(t) \propto t$ -- similar to the one-dimensional random walk problem \cite{wang45}. In order to keep the noise in the final results low, fast, non-adiabatic transformations are therefore favourable.

\subsection{Two-dimensional parametric space}

Although the one-dimensional spin transformation scheme can be implemented in a controllable manner and also the driving noise level optimised by a suitable driving, an important drawback
is the fact that after the transformation is completed the electron is spatially shifted from its initial position. 
This problem is resolved if the position of the quantum dot and the Rashba interaction are both time dependent, thus spanning a two-dimensional (2D) parameter space. As demonstrated in Ref.~\cite{cadez14}, the quantum dot can, for example, be first spatially shifted with some initial Rashba coupling value $\alpha_1$ and then displaced back to the starting position, while keeping the Rashba coupling fixed at different value $\alpha_2$ and finally setting the Rashba coupling back to its initial value $\alpha_1$. This transformation depends only on the area of the loop in the 2D parameter space. In particular, when the system is driven by a cyclic evolution, that is $\xi(t+T)=\xi(t)$, $\alpha(t+T)=\alpha(t)$,
$x_{c}(t+T)=x_{c}(t)$, $\dot{x}_{c}(t+T)=\dot{x}_{c}(t)$, $a_{c}(t+T)=a_{c}(t)$
and $\dot{a}_{c}(t+T)=\dot{a}_{c}(t)$, the angle of the spin
rotation around direction $\mathbf{n}$ is given by
\begin{equation}\label{phiT}
\phi(T)=\phi_A(T)=-2m^{*}\int_{0}^{T}\dot{a}_{c}(t)\xi(t){\rm d}t=2m^{*}\oint_{\mathcal{C}_{1}}a_{c}[\xi]{\rm d}\xi,
\end{equation}
where $a_{c}[\xi]$ represents the contour $\mathcal{C}_{1}$ in 2D parametric space $[\xi(t),a_{c}(t)]$ for $0\leq t\leq T$ thus
the spin rotation angle is  simply given by the area enclosed by 
$\mathcal{C}_{1}$.

Additionally to coordinate noises in 1D spin transformations, 
in 2D spin transformations are also normally distributed noise in initial SOI 
response $a_{c}(0)$, initial time derivative of the
response $\dot{a}_{c}(0)$ and stochastic noise in SOI
driving function  $\alpha(t)=\alpha^{0}(t)+\delta\alpha(t)$, where $\alpha^{0}(t)$ is ideal noiseless driving. SOI noise $\delta\alpha(t)$ is similar to the previous case of spatial driving and is again of the Ornstein-Uhlenbeck type with autocorrelation function $\langle\delta\alpha(t')\delta\alpha(t'')\rangle$, noise intensity $\sigma^2_\alpha$ and correlation time $\tau_\alpha$.

\begin{figure}[h]
	\centerline{\includegraphics[width=450pt]{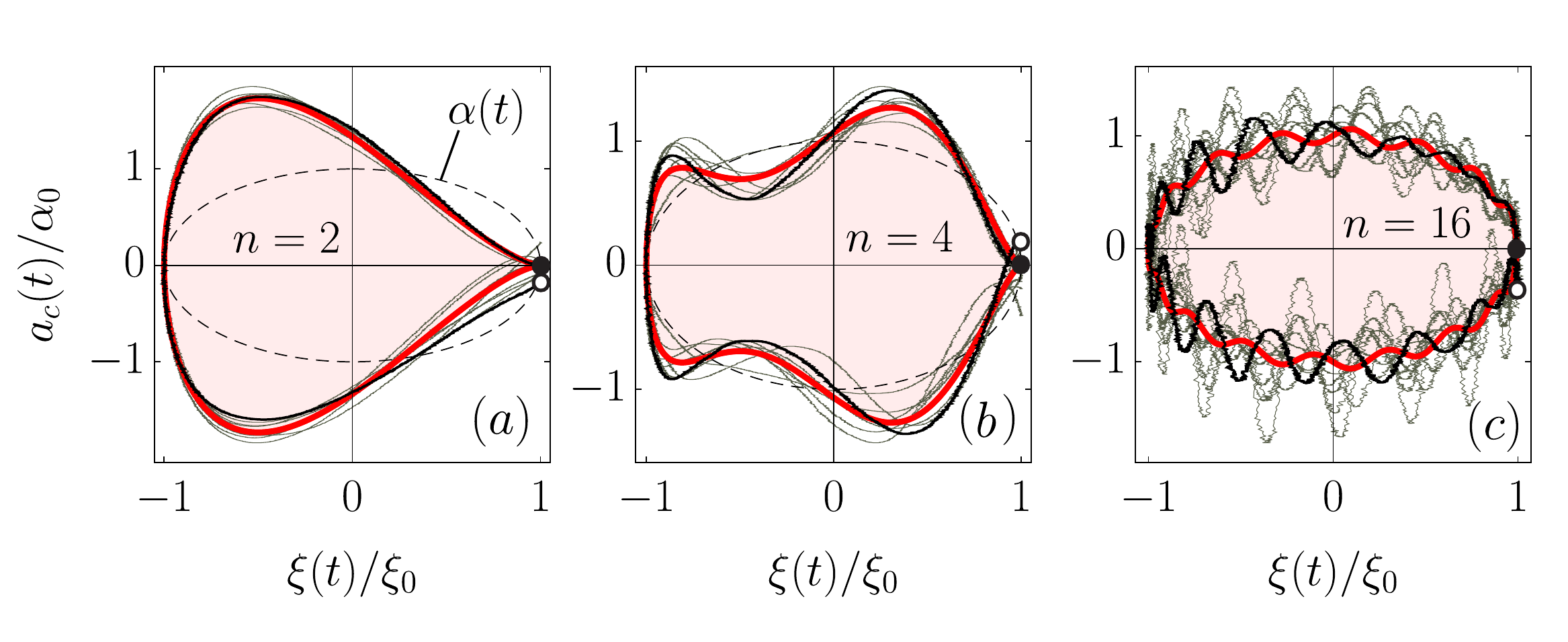}}
	\caption{Contours  ${\cal C}_1$ corresponding to equation~\eqref{eq:driving}, {\it i.e.}, $[\xi(t),a_{c}(t)]$  for $0 \leq t \leq T_n$  for  $n=2$ (a), $4$ (b) and $16$ (c) without noise (red lines) and 10 examples of results with superimposed white noise with $\omega \sigma^2_\xi/\xi^2_0=\omega \sigma^2_\alpha/\alpha^2_0=\frac{1}{400}$ (thin lines). Thick black lines show typical results, starting at positions $[\xi(0),\alpha(0)]$  (labelled by bullets) and ending at $[\delta\xi(T),\delta a_c(T)]$ (circles). Note that coloured area enclosed by ${\cal C}_1$ is proportional to the angle of spin rotation and that the limit $n\to \infty$ corresponds to the adiabatic regime of driving where ${\cal C}_1$ progressively approaches $[\xi(t),\alpha(t)]$ (dashed lines).}
	\label{fig:Driving}
\end{figure}

As an example of a 2D qubit rotation  we present the family of transformations with sinusoidal driving parametrized as
\begin{equation}\label{eq:driving}
\xi(t)=\xi_0 \sin(2 \pi t/T_n),\qquad \alpha(t)=\alpha_0 \cos(2 \pi t/T_n),
\end{equation}
with transformation times $T_n=   n T_1$, where $T_1=2\pi/\omega$ is the period of the confining potential and $n>1$. In figure \ref{fig:Driving} are shown some paths in the parametric
space $[\xi(t),a_{c}(t)]$ during the 2D transformation $0\leq t \leq T_n$. Thin lines represent ten different results with white noise. Time dependent variance $\sigma_a^2(t)$ is manifested as a spread of these curves around the ideal closed line ${\cal C}_1$ (red). Thick black lines represent typical case starting at positions $[\xi(0),\alpha(0)]$ (bullets) and ending at $[\delta\xi(T),\delta a_c(T)]$ (circles). It should be noted that accumulated errors in $a_c(T)=\delta a_c(T)$, with variance $\sigma_a^2$, are much larger than the corresponding $\delta\xi(T)$ which has only an instant white noise contribution. Finally, it should be noted also that the transformation angle $\phi$ -- proportional to the coloured area enclosed by ${\cal C}_1$ -- is due to oscillations of individual noisy curves around the ideal value relatively less prone to the noise.

The fact that, in comparison to 1D spin transformations,
these transformations include noise in
two dimensional parameter space,
where transformations need to be periodic, already indicates 
that fundamental differences might arise.
One such difference is in the effect of noise of initial conditions.
As seen from equation~\eqref{eq:harmoicsolution}, eventual nonzero initial conditions
directly contribute to response with cosine and sine time-dependency.
Spin transformations in 2D fulfil the condition of periodic response
$x_{c}(t+T)=x_{c}(t)$ and $a_{c}(t+T)=a_{c}(t)$ and 
have transformation times $T$ a multiple of the oscillator period $2\pi/\omega$. The errors in initial conditions then only translate the curve $\mathcal{C}_{1}$
in the parametric space but do not change the area enclosed by the contours and
thus the angle of rotation is not affected.

The only relevant source of gate noise is thus the noise in driving functions $\xi(t)$ and $\alpha(t)$, which induce noise 
in the appropriate variables which are for the case of white noise
all independent, stochastic and normally distributed
with the corresponding variances \cite{sed11}. As before, only short correlation times are considered with approximation, $\tau_\xi, \tau_\alpha \to 0$, leading to the time-dependent variances, $\sigma_{x}^{2}(t)$ as in equation \eqref{1d}, 
\begin{equation}\label{eq:sigmaxpc}
  \sigma_{\dot{x}}^{2}(t)=\frac{1}{4}\omega^{3}\sigma_{\xi}^{2}\left[2\omega t+\sin(2\omega t)\right],
\end{equation}
\begin{equation}\label{eq:sigmaac}
\sigma_{a}^{2}(t)=           {  \sigma_\alpha^{2} \over  \sigma_{\xi}^{2}} \sigma_{x}^{2}(t)  \quad {\rm and} \quad 
\sigma_{\dot{a}}^{2}(t)=         {  \sigma_\alpha^{2} \over  \sigma_{\xi}^{2}} \sigma_{\dot{x}}^{2}(t),  
\end{equation}
corresponding to  $x_{c}(t)$, 
$\dot{x}_{c}(t)$, $a_{c}(t)$ and $\dot{a}_{c}(t)$, respectively.
Induced noise in the angle of spin qubit rotation $\phi$ is then given by
\begin{equation}
\delta\phi=-2m^{*}\intop_{0}^{T}\dot{a}_{c}^{0}(t)\delta\xi(t){\rm d}t-2m^{*}\intop_{0}^{T}\delta\dot{a}_{c}(t)\xi^{0}(t){\rm d}t-2m^{*}\intop_{0}^{T}\delta\dot{a}_{c}(t)\delta\xi(t){\rm d}t,
\end{equation}
where $\dot{a}_c(t)=\dot{a}_c^{0}(t)+\delta\dot{a}_c(t)$,
$\dot{a}_c^{0}(t)$ being response without noise.
$\delta\phi$ is the sum of three independent and
normally distributed stochastic processes which lead to the variance 
$\sigma_{\phi}^2=\sigma_{I}^{2}+\sigma_{II}^{2}+\sigma_{III}^{2}$, 
where $\sigma_{I}^{2}$ is the variance of the first, $\sigma_{II}^{2}$
of the second and $\sigma_{III}^{2}$ of the third term \cite{feller71}. These terms are evaluated
directly from the phase average, for example 
$\sigma_{I}^{2}=4m^{*2}\langle \intop_{0}^{T}\dot{a}_{c}^{0}(t')\delta\xi(t')d t'
 \intop_{0}^{T}\dot{a}_{c}^{0}(t'')\delta\xi(t'')d t''\rangle$.
For arbitrary driving these terms are equal to
\begin{eqnarray}\label{gamma}
\sigma_{I}^{2}&=&(2m^{*}\sigma_{\xi})^{2}\intop_{0}^{T}\dot{a}_{c}^{0}(t)^{2}{\rm d} t=(2m^{*}\sigma_{\xi})^{2}\oint_{\mathcal{C}_{a}}\dot{a}_{c}^{0}[a_c^0]{\rm d}a_c^0,\\ \nonumber
\sigma_{II}^{2}&=&m^{*2}\omega^{3}\sigma_{\alpha}^{2}\intop_{0}^{T}\xi^0(t')\intop_{0}^{T}\{\sin\left[\omega\left(2\min(t',t'')-t'-t''\right)\right]\\
& &+\sin\left[\omega\left(t'+t''\right)\right]+2\omega\min(t',t'')\cos\left[\omega\left(t'-t''\right)\right]\}\xi^0(t''){\rm d} t''{\rm d} t',\\
\sigma_{III}^{2}&=&\frac{1}{4}\left(m^{*}\omega\sigma_{\xi}\sigma_{\alpha}\right)^{2}\left[\left(2\omega T\right)^{2}+2\sin^{2}\left(\omega T\right)\right].
\end{eqnarray}
The first contribution $\sigma^2_I$ is  proportional to the intensity of the $\xi$-noise and to the action integral associated with the SOI response, which vanishes in the adiabatic limit of the Rashba-driving. The second term, $\sigma^2_{II}$, originates in the $\alpha$-noise and is non-trivially related to the time dependence of the spatial driving function $\xi^0(t)$. As shown later in an example, this term can be made small by appropriate choice of driving. The last contribution  to the angle variance, $\sigma_{III}^{2}$, is of higher order in position and Rashba driving noise intensities and thus negligible for fast, non-adiabatic qubit  transformations while quadratically increasing for large $\omega T \gg 1$ adiabatic-like spin transformations. Variance $\sigma_{\phi}^2$ of noise in the angle of qubit rotation is therefore due to terms $I$ and $II$ enhanced for fast non-adiabatic drivings whereas it increases for large driving times due to the term $III$. This sets the condition for  minimal total induced noise. 

In order to elucidate this point we investigate angle variance for the circular driving scheme given by equation~\eqref{eq:driving}. At completion of the transformation at time $T_n$ the variance is given by
\begin{equation}\label{sigma0}
\frac{\sigma_{\phi,n}^{2}}{\phi_0^2}= \frac{n(1+n^{2})}{\pi(n^{2}-1)^{2}}\frac{\omega\sigma_{\xi}^2}{ \xi_0^2}  +
\frac{2 n^{3}}{\pi(n^{2}-1)^{2}}\frac{\omega\sigma_{\alpha}^2}{ \alpha_0^2}+
n^2\left(\frac{\omega\sigma_{\xi}\sigma_{\alpha}}{\xi_0\alpha_0}\right)^2,
\end{equation}
where $\phi_0=2\pi m^{*}\xi_0\alpha_0$  is the qubit rotation  angle equation~{\eqref{phiT} in the noiseless and $n\to\infty$ limit. 

In figure~\ref{fig:sigma}(a) is presented $\sigma_I$ as a function of time for various driving times for $n=2,4,8$ and $16$ with corresponding contours $[a_c^0(t),\dot{a}_{c}^{0}(t)]$ in figure~\ref{fig:sigma}(b). Note that at larger $n$ the area within particular contour $\mathcal{C}_{a}$ is progressively smaller -- as expected in the adiabatic limit. Figure~\ref{fig:sigma}(c) shows that the contribution $\sigma_{II}$ exhibits oscillations with time, but at final times the level of noise is minimum, decreasing with increasing $n$. The first two contributions to $\sigma_\phi^2$ are thus larger at short transformation times, {\it i.e.}, 
both decrease as $\propto 1/n$ for large $n\gg 1$ while the third term $\sigma_{III}^2$ increases as $\propto n^2$, figure~\ref{fig:sigma}(d), therefore there exists some minimum variance $\sigma_{\phi, n_{\rm min}}$ at optimal driving time for $n_{\rm min}$. For large $n\gg1$ such optimal regime can readily be evaluated,
\begin{equation}\label{eq:nmin}
n_{\rm min}=\left(\frac{\xi_{0}^2}{\pi\omega\sigma_{\xi}^2}+\frac{\alpha_{0}^2}{2\pi\omega\sigma_{\alpha}^{2}}\right)^{\frac{1}{3}},
\end{equation}
\begin{equation}\label{eq:sigmaPHImin}
\frac{\sigma_{\phi,n_{\rm min}}^{2}}{\phi_0^2}=3 n_{\mathrm{min}}^2 \left(\frac{\omega\sigma_{\xi}\sigma_{\alpha}}{\xi_0\alpha_0}\right)^2  \propto T_{n_{\rm min}}^2,
\end{equation}
therefore in order to minimize the variance the parameters should be chosen such that the driving time $T_{n_{\rm min}}$ is minimal.   The variance is for $T<T_{n_{\rm min}}$ limited by the extreme non-adiabatic value at $n=2$, 
\begin{equation}
\frac{\sigma_{\phi,n}^{2}}{\phi_0^2} < \frac{\omega\sigma_{\xi}^2}{\xi_0^2}  +
\frac{\omega\sigma_{\alpha}^2}{\alpha_0^2},
\end{equation}
which is qualitatively correct also for other types of driving with the contour ${\cal C}_1$ approximately bounded by the area $\xi_0 \alpha_0$, with noise intensities $\sigma_{\xi}$ and $\sigma_{\alpha}$.

\begin{figure}[h]
	\centerline{\includegraphics[width=330pt]{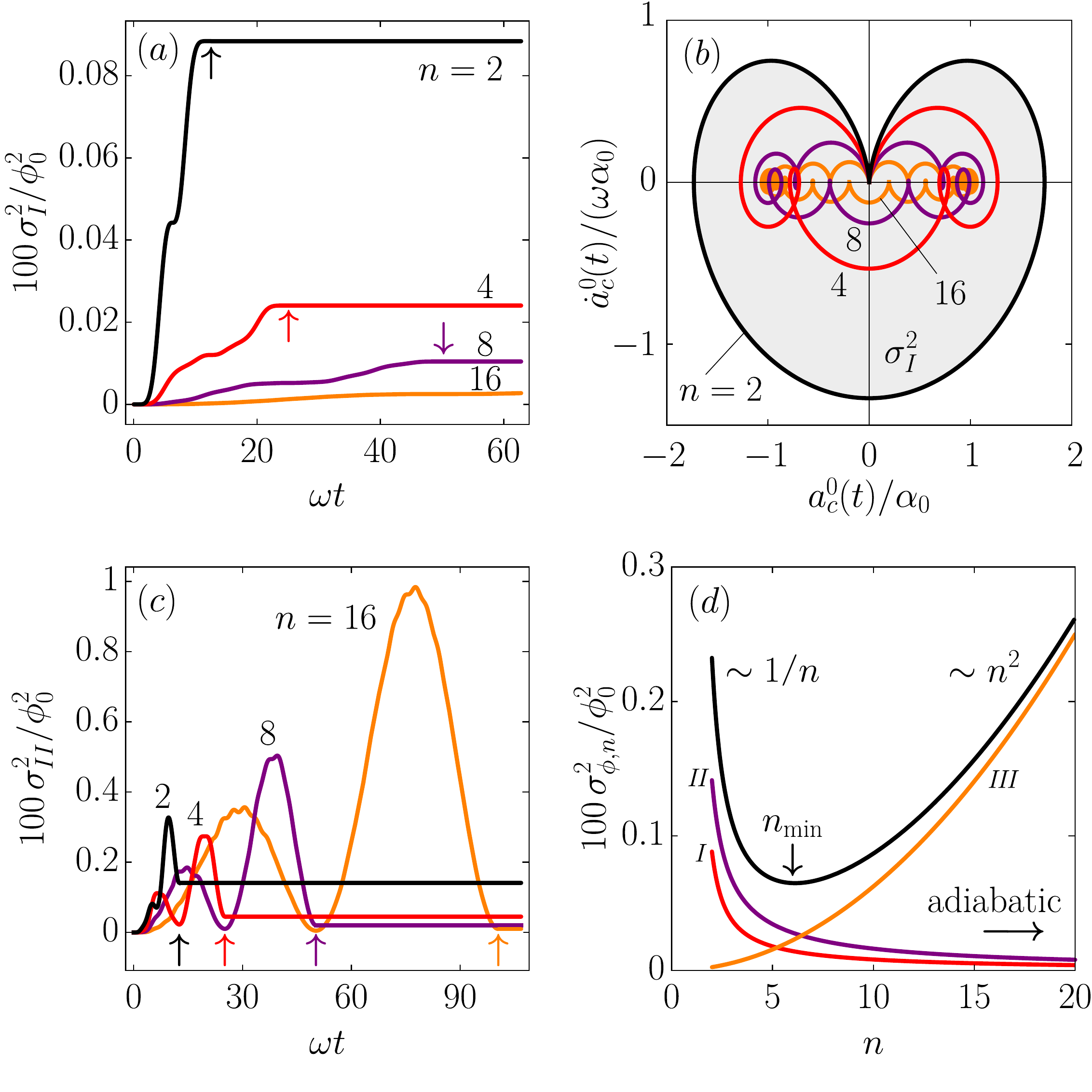}}
	\caption{Graphs are shown for different $n$ of sinusoidal driving, equation~\eqref{eq:driving}, at $\omega \sigma_{\xi}^2 / \xi_{0}^2 = \omega \sigma_{\alpha}^2 / \alpha_{0}^2= \frac{{1}}{400}$; scaled variances (a)  $\sigma_{I}^{2}(t)$ and (c) $\sigma_{II}^{2}(t)$, with coloured arrows pointing to final time $T=T_n$ of corresponding drivings (colours of arrows and drivings match). Phase space  contours $\mathcal{C}_{a}$, $[a_{c}^0(t),\dot{a}_{c}^0(t)]$, are shown in (b); note that gray shaded area is proportional to $\sigma_{I}^{2}(T)$ (for $n=2$, equation \eqref{gamma}]. In (d) is shown total variance $\sigma_{\phi,n}^{2}$  (black), together with variances $\sigma^2_{I,II,III}$ (red, violet and orange, respectively).}
	\label{fig:sigma}
\end{figure}

\section{Fidelity}

A fundamental property of adiabatic quantum phase is its invariance to changes in time-dependent Hamiltonian parameters, the actual phase being given by the area enclosed by the path in the parametric space. For fast, non-adiabatic holonomic transformations considered here, the phase  is given by the area, in combined space of driving and response parameters, which only in the adiabatic limit is independent of time. More importantly, for non-adiabatic qubit manipulations the electron state is determined by the time-dependent Hamiltonian during the evolution and will in general be a superposition of excited states,
becoming the ground state when the transformation is complete. As shown in Refs.\cite{cadez13,cadez14,kregar16} such motions in parametric space can easily be performed if the driving functions are appropriately chosen. 

In the previous Section the  analysis of spin-rotation angle variance demonstrated that  due to gate noise in the driving functions, spin transformations are not completely faithful and that additional fine tuning is required in order to minimise these noise effects. Moreover, for the present case of non-adiabatic qubit transformations, an important additional question is relevant: how well does the final state of the electron relax to the desired final state energy manifold after the transformation if the driving function is not ideal as in the presence of noise?  

In order to answer this question we consider the qubit wave function  $|\Psi_{0\frac{1}{2}}(t)\rangle$, equation~\eqref{psi}, at $t=0$ in the ground state of the harmonic quantum dot  with $m=0$ and spin $\frac{1}{2}$. We observe the relaxation to the ground state, spanned by the basis of the time dependent Hamiltonian equation~\eqref{H}  at time $t$ \cite{cadez13},
\begin{equation}
|\widetilde\Psi_{0s}\rangle=e^{-im^{*}[x-\xi(t)]\alpha(t)\mathbf{n}\boldsymbol{\cdot\sigma}}|\psi_{0}[x-\xi(t)]\rangle|\chi_{s}\rangle.
\end{equation}
As the appropriate measure of the relaxation accuracy we define fidelity $F=\langle\Psi_{0\frac{1}{2}}(t)|P_0|\Psi_{0\frac{1}{2}}(t)\rangle$, where  $P_0=\sum_{s}|\widetilde\Psi_{0s}\rangle\langle\widetilde\Psi_{0s}|$ is the projector onto the ground state manifold.  We choose $\mathbf{n}$ perpendicular to the $z$-axis and a lengthy but straightforward derivation yields 
the expression 
\begin{equation}
F=\frac{1}{2}(e^{-E_{+}}+e^{-E_{-}}),
\end{equation}
\begin{equation}
E_{\pm}=\frac{m^{*}}{2\omega}\{[\omega(x_{c}(t)-\xi(t))\pm\dot{a}_{c}(t)/\omega]^{2}+[\dot{x}_{c}(t)\mp(a_{c}(t)-\alpha(t))]^{2}\},
\end{equation}
where $E_{\pm}$ resembles classical energy
with additional terms for spin-orbit coupling and is equal to the classical
energy if the spin-orbit driving is constant \cite{cadez13}. Fidelity for 1D driving is obtained
as a limit of 2D case when one of the drivings is constant,
for example $a_{c}(t)=\alpha_0$, $\dot{a}_c(t)=0$ 
thus $\sigma_{\alpha}=\sigma_{a_{c}}=\sigma_{\dot{a}_{c}}=0$.
The expression for fidelity then simplifies to 
$F=e^{-E}$, where $E=E_{+}=E_{-}$ is now equal to the classical energy of 
harmonic oscillator. 

Ideal qubit transformations, $F=1$, are achieved  by applying ideal drivings, where the energies $E_\pm$ vanish at final time $t=T$, {\it i.e.}, when $x_c=\xi$, $a_c=\alpha$,  $\dot{x}_{c}=0$, and $\dot{a}_{c}=0$. Hovewer,
the presence of noise in spin-orbit and spatial driving terms makes fidelity a random quantity, 
described by a probability density function $\frac{\mathrm{d}P(F)}{\mathrm{d}F}$. It 
can be calculated from the probability density for variables $E_{\pm}=f_{\pm}(x_{c},\dot{x}_{c},a_{c},\dot{a}_c)$ which  
are functions of independent random variables, normally distributed and with variances equations
\eqref{eq:sigmaxpc}{--}\eqref{eq:sigmaac}.
The probability density functions for  $E_{\pm}$  can be calculated using the formula 
\begin{eqnarray}
\frac{\mathrm{d}P(E_{\pm})}{\mathrm{d}E_{\pm}}&=&\int_{-\infty}^{\infty}\!\!\!\!\!\dots\int_{-\infty}^{\infty}
\frac{\mathrm{d}P(x_{c})}{\mathrm{d}x_{c}}\frac{\mathrm{d}P(\dot{x}_{c})}{\mathrm{d}\dot{x}_{c}}
\frac{\mathrm{d}P(a_{c})}{\mathrm{d}a_{c}}\frac{\mathrm{d}P(\dot{a}_{c})}{\mathrm{d}\dot{a}_{c}}
\nonumber
\\
& &\times
\delta[E_{\pm}-f_\pm(x_{c},\dot{x}_{c},a_{c},\dot{a}_{c})]
\mathrm{d}x_{c}\mathrm{d}\dot{x}_{c}\mathrm{d}a_{c}\mathrm{d}\dot{a}_{c}.
\end{eqnarray}
The result is obtained by first calculating the characteristic function, 
followed by the inverse Fourier transform yielding distributions with the same functional form for variables $E_{+}$ and $E_{-}$, 
\begin{equation}
\frac{\mathrm{d}P(E_{\pm})}{\mathrm{d}E_{\pm}}=2\sigma_{1}^{-1}\sigma_{2}^{-1}I_{0}[(\sigma_{1}^{-2}-\sigma_{2}^{-2})E_{\pm}]e^{-(\sigma_{1}^{-2}+\sigma_{2}^{-2})E_{\pm}},
\end{equation}
with
\begin{equation}
\sigma_{1}^{2}(t)=\left(\frac{2m^{*}}{\omega}\right)
(\omega^{2}\sigma_{x}^{2}(t)+\sigma_{\dot{a}}^{2}(t)/\omega^{2}),
\end{equation}
\begin{equation}
\sigma_{2}^{2}(t)=\left(\frac{2m^{*}}{\omega}\right)
\left(\sigma_{\dot{x}}^{2}(t)+\sigma_{a}^{2}(t)\right),
\end{equation}
where $I_{0}(z)$ is the modified Bessel function of the first kind.

Distributions for $e^{-E_{\pm}}$ are calculated by using 
a simple change of variables formula. Since the
fidelity is a sum of two dependent random variables, its 
probability distribution is calculated from the joint probability distribution 
function for those two variables, which in general cannot be evaluated analytically. However, one can examine $\frac{\mathrm{d}P}{\mathrm{d}F}$ exactly
in two convenient limiting  cases. The first is the case when the noise in one of the driving variables, for example 
$\alpha(t)$, is much weaker than the other, {\it i.e.}, $\sigma_{\alpha}\ll\sigma_{\xi}$. In this case, the noise properties are essentially those of a 1D problem and the exact expression for probability density function 
of fidelity is 
\begin{equation}\label{eq:fidelity1d}
\frac{\mathrm{d}P(F)}{\mathrm{d}F}=2\sigma_{1}^{-1}\sigma_{2}^{-1}I_{0}[(\sigma_{1}^{-2}-\sigma_{2}^{-2})\ln F]F^{\sigma_{F}^{-2}-1}, \quad 
\sigma_{F}^{-2}=\sigma_{1}^{-2}+\sigma_{2}^{-2},
\end{equation}
which  for $\sigma_{1,2}\to0$ leads to $\frac{\mathrm{d}P}{\mathrm{d}F} \propto F^{\sigma_{F}^{-2}}$.

The second limiting case is when $\sigma_{x}^{2}(t)=\sigma_{\dot{a}}^{2}(t)/\omega^{4}$ and $\sigma_{\dot{x}}^{2}(t)=\sigma_{a}^{2}(t)$, which is satisfied for $t=T_n$ if  the coordinate and the SOI driving noise intensities are equal, {\it i.e.}, 
$\sigma_\alpha=\omega\sigma_{\xi}$. 
In this case $E_{+}$ and $E_{-}$ are
independent random variables \cite{feller71} and $\frac{\mathrm{d}P}{\mathrm{d}F}$
can be calculated as the convolution of probability distributions for
$e^{-E_{+}}$ and $e^{-E_{-}}$. The exact result for $F\geq\frac{1}{2}$ is
\begin{equation}\label{eq:fidelity2d}
\frac{\mathrm{d}P(F)}{\mathrm{d}F}=2\sigma_{F}^{-4}[B(\frac{1}{2F},\sigma_{F}^{-2},\sigma_{F}^{-2})-B(1-\frac{1}{2F},\sigma_{F}^{-2},\sigma_{F}^{-2})](2F)^{2\sigma_{F}^{-2}-1},
\end{equation}
where $B(x,a,b)$ is the incomplete beta function. For $F <\frac{1}{2}$ the probality distribution is given by
$\frac{\mathrm{d}P}{\mathrm{d}F}=2\sigma_{F}^{-4}B(\sigma_{F}^{-2},\sigma_{F}^{-2})(2F)^{2\sigma_{F}^{-2}-1}$, where $B(a,b)$ is the beta function.

In practice the most relevant regime is $\sigma_{F}\to0$ for which the probability distribution equation~(\ref{eq:fidelity2d})  simplifies to  $\frac{\mathrm{d}P}{\mathrm{d}F} \propto (1-F)F^{2\sigma_{F}^{-2}}$.
\begin{figure}[h]
	\includegraphics[width=450pt]{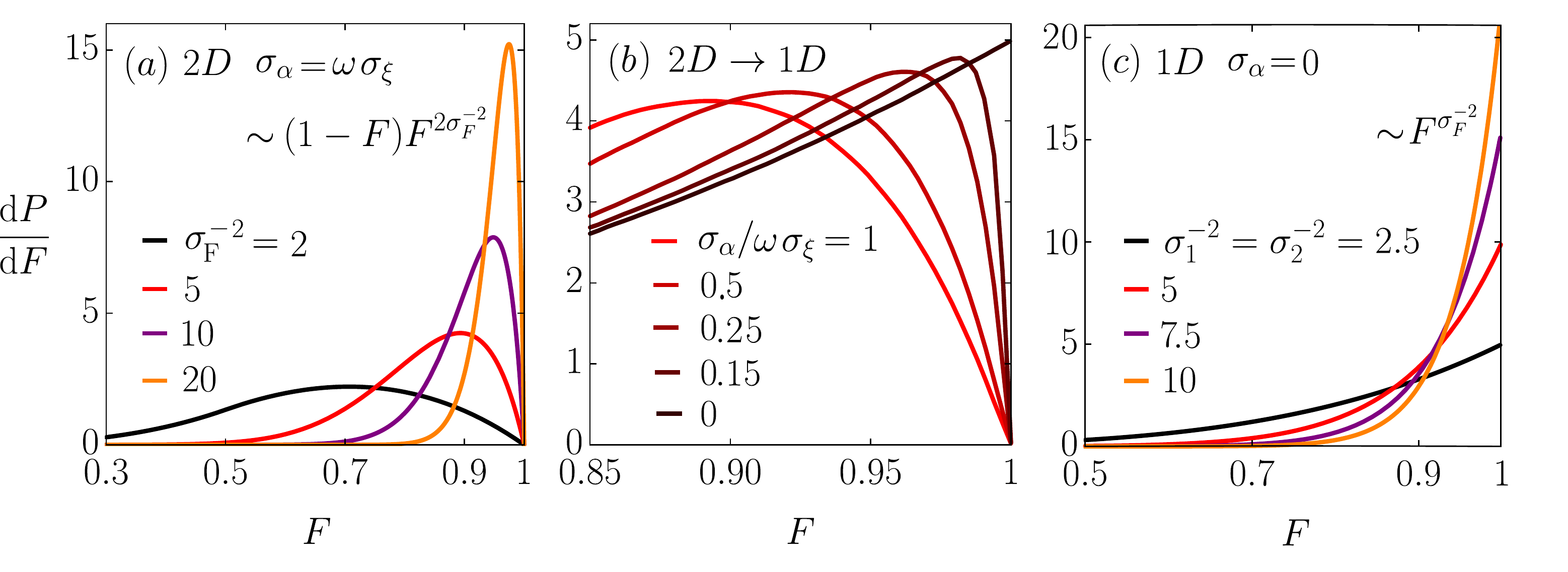}
	\caption{(a) Probability density function of fidelity for 2D  [$\sigma_{\alpha}=\omega\sigma_\xi$, equation \eqref{eq:fidelity2d}] and (c) 1D  
		transformations is shown for different values of noise [$\sigma_{\alpha}=0$, equation \eqref{eq:fidelity1d}]. (b) Numerical results for 2D probability density of fidelity with different ratios of noise intensity in coordinate and SOI driving. As ratio decreases from $\sigma_{\alpha}/\omega\sigma_\xi=1$ (red), probability density function of fidelity transforms to the form of 1D transformations (black). Parameters $\sigma_\xi$ and $\sigma_\alpha$ are such that $\sigma_F^{-2}=5$ at all ratios.}
	\label{fig:Fidelity}
\end{figure}
At  first glance the surprising result is that 
the probability distribution for $F\to 1$ tends to zero, 
in other words, it is not possible to exactly  achieve a flawless spin-flip
transformation with 2D driving. This is quite different from transformations with 1D driving, where the probability for flawless
transformation at $F=1$ is maximum. For intuitive interpretation one can 
compare the noise in 1D and 2D transformations to Brownian motion in 1D and 2D parameter space
and the fidelity to the probability of finding a particle after some time near the starting point \cite{r05,meinrichs93}. In 1D the particle always returns to the starting point while in 2D the particle returns {\it almost surely} with maximum probability at the annulus near the starting point. 

In figure \ref{fig:Fidelity}(a) is shown the fidelity probability density function equation~(\ref{eq:fidelity2d})
for different $\sigma_{F}$ . It is clear that although the probability for driving flawlessly is zero, the 
position of the maximum of $\frac{\mathrm{d}P}{\mathrm{d}F}$ is for small $\sigma_F$ very close to 1,
\begin{equation}
F_{\rm max}=1-\frac{1}{2}\sigma_{F}^{2},
\end{equation}
the width at half maximum is
\begin{equation}
\sigma_{2{\rm D}}=\frac{e}{2}\,\sigma_{F}^{2},
\end{equation}
and linear growth of probability from zero to maximum value indicates that when the 
noise is low enough, the transformation is with very high probability achieved
with almost zero error. Note that the fidelity is less sensitive to the noise than the transformation angle studied in Section 3 as indicated by the quadratic dependence of the shift in fidelity distribution maximum away from $F=1$.
Hence $\delta F\sim\sigma_{F}^{2}$ is of higher order in the noise intensities than the width of the transformation angle,  $\sigma_{\phi}\sim\sigma_\xi$ (or $ \sigma_\alpha$).

Similar arguments apply to the qubit transformations with 1D driving,
where the position of the maximum of $\frac{\mathrm{d}P}{\mathrm{d}F}$ is equal to $1$
and full width at half maximum $\sigma_{1{\rm D}}$ is given by
\begin{equation}
\sigma_{1{\rm D}}=\mathrm{ln}2\,\sigma_{F}^2.
\end{equation}
Examples of numerically generated results for general $\sigma_\xi$ and $\sigma_\alpha$ \cite{mp89} are shown in figure \ref{fig:Fidelity}(b), where
in the limit $\sigma_\alpha/\sigma_\xi\to 0$ the probability distribution gradually transforms from 2D to 1D form and in figure \ref{fig:Fidelity}(c) such
1D results are presented for various $\sigma_1=\sigma_2$.

For the example of sinusoidal driving, equation~ \eqref{eq:driving} considered in the previous Section, the lowest noise in the angle of
spin rotation is achieved for driving times $T_{n_{\rm min}}=n_{\rm min}T_1$, equation~\eqref{eq:nmin}. The corresponding
probability distribution of fidelity is equal to equation~\eqref{eq:fidelity2d}, with 
\begin{equation}
\sigma_{F}^{2}= \frac{1}{2} m^{*}(\omega^2\sigma_{\xi}^{2}+\sigma_{\alpha}^{2})\omega T_{n_{\rm min}},
\end{equation}
therefore regarding the minimisation of  both variances, for rotation angle and for fidelity, faster transformations with lower $T_{n_{\rm min}}$ are favourable.

\section{Summary and conclusion}

Recent theoretical analysis has revealed that holonomic spin manipulation in linear systems \cite{cadez13,cadez14} or on an appropriate ring support \cite{kregar16,kregar16b} is feasible from adiabatic to strong non-adiabatic regime of driving. The first prerequisite here is the ability to control the position of the electron $\xi(t)$ and the second is controllable manipulation of the Rashba coupling, regarding the time dependent strength $\alpha(t)$ and also the choice of preferred  direction ${\bf n}$. For slow, adiabatic qubit manipulation these requirements lead to an arbitrary transformation, simply determined by the area in the space of driving parameters $[\xi,\alpha]$. During the process of the transformation the electron remains permanently in the same spatial state, the ground state for example, and only spin properties change.

Fast, non-adiabatic spin manipulation is far more challenging since the time-dependence of driving functions  have to be appropriately tuned. Unlike the adiabatic regime, the transformation angle of spin is given by the combined space of both the driving function $\xi(t)$ and the SOI response $a_c(t)$ to the driving function $\alpha(t)$. In addition to correct transformation of the spin direction, one has also to take care that the electron state has not left the starting energy manifold at the final time. For example, starting from the ground state the electron should, after performing one cycle with time-dependent Hamiltonian, return to the ground state, although during the cycle the state of the electron may be a superposition of excited eigenstates of the moving potentials. As shown in Refs.\cite{cadez13,cadez14,kregar16} such drivings are feasible to perform. As long as the approximation of the harmonic potential is justified, the formalism yields exact time dependent wave functions  with simple tuning of driving functions in order to achieve desired qubit transformations. However, noise in driving functions will always be present because of unavoidable gate noise, which means that qubit transformation will always deviate from the ideal one.

In this paper we examined in detail the influence on qubit transformations of various imperfections in driving. The formalism allows analytical treatment of arbitrary driving, therefore we concentrated on the exact analysis of the influence of small deviations from ideal qubit manipulation. In particular, for 1D manipulation we show how errors in initial conditions give rise to variance in the transformation angle. It is shown how one can analyse the effects of a general coloured noise to the transformation angle and, as an example, we show the result for Ornstein-Uhlenbeck noise in the limit of short correlation times (white noise) although the formalism can be applied to other types of noise defined by their autocorrelation functions.

The results valid for 1D parametric space are generalised to more involved analysis of the transformation angle for the case of 2D spin manipulation with time dependence of both quantum dot position and SOI. The first result here is that, due to periodicity, holonomic manipulation is completely insensitive to the initial conditions, since the qubit rotation angle is given solely by the area in parameter space for which errors in the starting point are irrelevant. Therefore the only source of errors here is the noise in driving functions. As in the 1D case exact results can be derived for a broad class of coloured noise with given autocorrelation functions and appropriate formulae are given explicitly. As a typical example, considered in detail, is the case of circular driving in the space of parameters for which exact analytical formulae are given and analysed for white noise.  It is argued that these particular results are qualitatively valid in general, providing similar size of the contour in parametric space and similar noise intensities. In particular,  for non-adiabatic manipulations, errors increase due to the detuning of sensitivity drivings in the presence of the noise, while in the adiabatic limit of driving the accumulation of errors is similar to random walk process. In general we expect some optimum regime between non-adiabatic and adiabatic driving and we also show that minimal variance can be achieved by suitable tuning. 

As discussed above, for non-adiabatic regimes a non-trivial point of issue is the ability of the system to return to the ground state after an arbitrary time-dependent driving. For that reason our analysis was focused on fidelity - the overlap of the actual wave function with the desired ideal. For the white noise limit of coloured noise explicit formulae are derived. For the 1D case and general time-dependent variances of response functions the result is given explicitly. For 2D, exact analytical results are derived for symmetric noise in position and spin-orbit driving functions. For more general cases some examples are calculated numerically and shown to demonstrate smooth transition between two limiting cases, totally symmetric 2D and asymmetric 1D. 

We conclude with an interesting observation that the noise effects on fidelity have a structure similar to probability density in random walk problems. In 1D random walk the particle always returns to the origin and similarly the fidelity probability distribution $\sim F^{\sigma_F^{-2}}$ for the 1D parametric case which exhibits a maximum at $F=1$, {\it i.e.}, although the qubit motion is influenced by the random noise, the wave function still returns with maximum probability to the ground state. In a 2D random walk the classic result \cite{r05} is that the particle after some elapsed time returns to the origin {\it almost surely}  but with maximum probability at annulus displaced from the origin. Similarly in our case the maximum of the fidelity probability distribution $\sim (1-F) F^{2\sigma_F^{-2}}$ is slightly shifted away from $F=1$. Finally, we show that errors in fidelity occur at a higher order in noise intensity compared with errors in qubit rotation angle.

\ack
The authors thank J H Jefferson,  A Kregar, T Rejec and S \v Sirca for valuable suggestions and acknowledge support from the Slovenian Research Agency under contract no. P1-0044.

\end{document}